\documentclass[12pt,preprint,longabstract]{aastex}

\shortauthors{Proffitt et al.}
\shorttitle{Limits on $\epsilon$~Eridani Dust }

\begin{document}

\title{Limits on the Optical Brightness of the $\epsilon$~Eridani Dust Ring$^1$
}

\altaffiltext1{Based on observations made with the
NASA/ESA Hubble Space Telescope, obtained at the Space Telescope
Science Institute, which is operated by the Association of
Universities for Research in Astronomy, Inc., under NASA contract NAS
5-26555. These observations are associated with proposal
GO-09037.}

\author{Charles R. Proffitt\altaffilmark{2}}
\affil{Science Programs, Computer Sciences Corporation,
3700 San Martin Drive, Baltimore, MD 21218;
proffitt@stsci.edu}

\author{Kailash Sahu, Mario Livio, John Krist, Daniela Calzetti, Ron Gilliland} 
\affil{Space Telescope Science Institute, Baltimore, MD 21218}

\author{Carol Grady\altaffilmark{3}} 
\affil{NOAO/STIS, Code 681, NASA Goddard Space Flight Center, Greenbelt, MD 20771}

\author{Don Lindler}
\affil{Sigma Research and Engineering, Lanham, MD 20706}

\author{Bruce Woodgate\altaffilmark{4}, Sara Heap\altaffilmark{4}, Mark Clampin \& 
Theodore R. Gull\altaffilmark{4}}
\affil{Code 681, Laboratory for Astronomy and Solar Physics, NASA Goddard Space Flight
Center, Greenbelt, MD 20771}

\author{Casey M. Lisse}
\affil{University of Maryland, Department of Astronomy, College Park, MD
20742\\ }

\altaffiltext2{also the Space Telescope Science Institute, and the Institute
for Astrophysics and Computational Science at the Catholic University of
America.}

\altaffiltext3{Also Eureka Scientific, 2452 Delmer Street, Suite 100, Oakland, CA
94602-3017}

\altaffiltext4{Member of Space Telescope Imaging Spectrograph Investigation
Definition Team}


\begin{abstract}

The STIS/CCD camera on the {\em Hubble Space Telescope (HST)} was used to take
deep optical images  near the K2V main-sequence star $\epsilon$~Eridani in an
attempt to find an optical counterpart of the dust ring previously imaged by
sub-mm observations. Upper limits for the optical brightness of the dust ring
are determined and discussed in the context of the scattered starlight expected
from plausible dust models. We find that, even if the dust is smoothly
distributed in symmetrical rings, the optical surface brightness of the dust,
as measured with the {\em HST}/STIS CCD clear aperture at 55 AU from the star,
cannot be brighter than about 25 STMAG/"$^2$. This upper limit excludes some
solid grain models for the dust ring that can fit the IR and sub-mm data.
Magnitudes and positions for $\approx\,$59 discrete objects between 12.5" to
58" from $\epsilon$~Eri are reported. Most if not all of these objects are
likely to be background stars and galaxies.

\end{abstract}


\section{Introduction.}

A substantial fraction ($\approx15$\%) of main-sequence stars show
evidence for excess IR or sub-mm flux due to thermal emission from dust
located at distances of 30 AU or more from the stars; i.e., locations 
comparable to that of the Kuiper Belt in our own Solar System. This was
first discovered using IRAS observations (e.g., Aumann et al.\ 1984; 1985;
Gillett \& Aumann 1983), and subsequently, observations with  the James
Clarke Maxwell Telescope's Submillimeter Common-User Bolometer Array
(JCMT/SCUBA) have directly imaged the dust distribution in a few of
these systems (Holland et al.\ 1998), including $\epsilon$~Eri (Greaves et
al 1998).

$\epsilon$~Eridani is a K2V main-sequence star at a distance of about
3.2 pc. It is believed to be a relatively young system ($< 1$ Gyr; Song
et al.\ 2000; Soderblom \& Dappen 1989), with a mass slightly less than
our own Sun. From radial velocity measurements, Hatzes et al (2000) have
reported evidence for a planet in this system with a semimajor axis of
3.4 AU and $m\sin i=0.86 M_J$.

The 850$\,\mu$m observations of  Greaves et al.\ show a ring-like structure
around $\epsilon$~Eri. The maximum surface brightness of this ring is
located at a radius of $\approx\,$17" (55 AU) from the star, with some flux
extending out as far as 36" (115 AU). The observed 850$\,\mu$m flux shows
the ring to be asymmetrical, with several bright clumps. It has been
suggested that structures of this kind can be caused by resonant interactions
of dust with planets in or near the ring (Liou \& Zook 1999; Ozernoy et al
2000; Quillen \& Thorndike 2002).

In an attempt to detect an optical counterpart of this ring we undertook
observations with the {\em Hubble Space Telescope's} Space Telescope
Imaging Spectrograph ({\em HST}/STIS) of $\epsilon$~Eri, using this
instrument's CCD camera, as part of HST GO program 9037 (Mario Livio PI).
A full description of the STIS instrument can be found in Kim Quijano et al.\ 
(2003).

While this camera can be used with a number of coronagraphic wedges,
saturation of the detector by the wings of the stellar PSF near the edges
of the wedge would severely limit the exposure time achievable in a single
image.  Therefore, we instead used the 52"X52" clear CCD aperture and
placed the star $\approx5$" off the edge of the detector. The K0~IV star
$\delta$~Eri was also observed as a PSF comparison star.

\subsection{Details of Previous Observations}

Gillett (1986) reconsidered the IRAS observations of $\epsilon$~Eri and
concluded that the intrinsic FWHM of the source flux at 60$\,\mu$m was
less than 17" in the IRAS scan direction and less than 11" in the
perpendicular direction. If this suggestion is correct, it would imply
that the bulk of the IRAS emission comes from a region inside of the
ring detected by Greaves et al.\ (1998).  However, both the sub-mm ring
and the IRAS size limits suggested by Gillett are significantly smaller
than the nominal IRAS 60$\,\mu$m resolution at of about 1' (Beichman et
al., 1985), and therefore Gillett's conclusions should be treated
cautiously.

Greaves et al.\ observed the system with SCUBA at both 850 and
450$\,\mu$m. The 850 $\,\mu$m observations used a beam of 15" FWHM, and
Greaves et al published smoothed versions of the images produced by
these observations. The S/N of the  450$\,\mu$m
observations is too low to give any useful spatial information (no image
was published), but it does supply a useful measure of the total flux.

Sch\"utz et al.\ (2004) obtained 1200 $\mu$m measurements using a 25"
beam, that are consistent with Greave et al.'s measurements. Several
previous sub-mm observations (Chini et al.\ 1990; 1991; Zuckerman \&
Becklin 1993) at various wavelengths (800 -- 1300 $\mu$m) had used
single pointings with  beam-sizes and background chopping too small to
properly measure the structure detected by Greaves et al.\ (see also
Weintraub \& Stern 1994). While these observations may provide some
useful constraints, they cannot be used as direct measures of the flux
and are not further considered here.

We summarize the IRAS,  Greaves et al., and Sch\"utz et al.\ data in
Table \ref{fluxtab}. In this table we adopt Greaves et al's color
corrections, as well as their corrections for the stellar contribution
to the total flux, and present only the flux attributed to the dust
alone.

\placetable{fluxtab}

At 55 AU, the IRAS and SCUBA data imply grain temperatures of about 30 K
-- close to the equilibrium black body temperature. The very flat
450$\,\mu$m to 850$\,\mu$m flux ratio ($4.6\pm2.6$) requires the
presence of large grains ($> 100\,\mu$m), which can emit efficiently at
sub-mm wavelengths. At this distance, the time scale for the orbital
decay of 100$\,\mu$m grains due to the Poynting-Robertson effect is
about $7\times10^8\,$yr; a time scale which is comparable to the
inferred age of the system. This would seem to argue against a
substantial population of smaller grains in the outer parts of the
system. However, recent works on various debris disk systems (e.g., Wyatt
et al.\ 1999; Li, Lunine, \& Bendo 2003) have shown that this kind of
interpretation is overly simplistic. The sub-mm emissivity of grains
drops rapidly enough with decreasing grain size that a substantial
population of smaller grains has little effect on the sub-mm flux
ratios, and the collision and fragmentation rate of the large grains
required by the sub-mm data is still high enough to
replenish the smaller grains faster than the Poynting-Robertson effect
can remove them. We will see that the nature and abundance of these
smaller grains has dramatic effects on the optical detectability of the
dust in the $\epsilon$~Eri system.

\section{Description of the Observations}

The HST observations in this program were done on January 26, 2002 using six
adjacent single-orbit visits (see Table \ref{obstable}). In each visit, after
taking an ACQ exposure to determine the position of the targeted star, 15
offset exposures of 109 seconds each were taken using the unfiltered STIS CCD 
in imaging mode.
During these offset exposures the bright star was
located about 5" off the detector (near CCD pixel coordinates x=1127, y=514).
The first and last orbits were used to observe the PSF of the comparison star,
$\delta$~Eri at two different orientations differing by 30$^\circ$. The four
intermediate orbits were used to observe the primary target at four different
orientations separated by 10$^\circ$ intervals. In each of these latter
exposures, the position of the brightest sub-millimeter clump observed by
Greaves et al.\ was imaged on the 1024x1024 pixel detector.

The clear aperture used with the STIS CCD is designated as the "50CCD" aperture
in STIS documentation. It has a field of view of almost 52"$\times$52", and has
a very broad bandpass, with significant throughput from about 2000 \AA\ to
10200 \AA. Prior to the installation of ACS, STIS 50CCD observations provided
the most sensitive {\em HST} mode for deep imaging. We will, for the most part,
give our observed STIS magnitudes in STMAG units, where the magnitude
is defined as $-2.5\log(F_\lambda)-21.10$, with $F_\lambda$ in
ergs$/$s$/$cm$^2$/\AA.  For STIS 50CCD imaging, the conversion to a magnitude
system that uses Vega as the zero point is VEGAMAG(50CCD)$=$STMAG(50CCD)$-0.36$.

\placetable{obstable}

In addition, at the beginning of each orbit a very short observation was
taken of each star using the STIS CCD with the F25ND3 filter
(Table~\ref{nd3images}). 

\placetable{nd3images}

\section{Data Analysis}

\subsection{Basic Analysis}

The standard STIS pipeline software was used to produce bias and dark
subtracted and flat fielded images of individual subexposures (flt files).
Comparison
with median filtered images was done to identify and correct hot pixels
that were not handled properly in the standard analysis; about 0.5\% of
pixels were corrected in this way.

Each of the 15 subexposure flt files was rearranged into five separate
files, each containing three adjacent subexposures, and these files
were input into the stsdas stis routine ocrreject to produce five
separate cosmic ray rejected (crj) files for each of the six visits - 30 crj files in total. 

In each of these crj files, the two diffraction spikes from the  star
($\epsilon$ Eri or $\delta$ Eri) located 5" off the edge of the detector are
the brightest features visible. These were used to register the images. As we
are primarily interested in the relative offsets between the images, any small
systematic offset from the real location of the star relative to the detector
is unimportant. Each of the 30 crj files was then shifted so as to put the
intersection of the diffraction spikes at the mean location measured for all
visits. The shifts applied were up to $\pm0.7$ pixels in x and $\pm0.3$ pixels
in y. Because we wanted to distort the high S/N pattern of the PSF as little as
possible, we shifted the images using a seven point sinc interpolation function
with the iraf imshift routine. The shifted crj files for each visit were then
combined by again using the STIS ocrreject routine,  producing a single aligned
crj file for each visit. 

We investigated whether there was any advantage in shifting each
individual subexposure, rather than combining them first by groups of
three before shifting and coadding, but this does not appear to
significantly improve the final  coadded image.

  \subsection{PSF subtraction}

Subtraction of the PSF from the wings of the a bright star is very
sensitive to small mismatches in the target and PSF star's spectral
energy distribution, as well as to small changes in telescope focus and
breathing (i.e., changes in image quality caused by flexure of {\em HST}
and STIS optical elements).  We will use two separate techniques to subtract
$\epsilon$~Eri's PSF.  Roll deconvolution techniques use the target as
its own PSF star, by taking back-to-back observations at different
orientations. Direct subtraction of the $\epsilon$~Eri and $\delta$~Eri
observations will also be done. A comparison of different PSF
subtraction techniques for STIS Coronagraphic observations was done by
Grady et al. (2003), and much of their discussion will also be relevant
here.

  \subsubsection{Roll Deconvolution of $\epsilon$ Eri}

The goal of the roll deconvolution is to use images taken at different
orientations to separate the real sky image from the PSF of the bright
nearby star. This technique eliminates any problem with mismatches in
the shape of the PSF due to differences in the spectral energy
distributions, but it has the disadvantage that any circularly symmetric
features or arc-like structures larger than the change in roll angle
will be included with the PSF rather than as part of the sky image.
Unfortunately these are exactly the type of structures most likely in a
circumstellar debris disk.

To obtain a first approximation to the PSF of $\epsilon$~Eri, we
combined the four aligned and coadded $\epsilon$~Eri crj images,
rejecting points that were high or low by more than three sigma from the
median value for that pixel location by using the iraf imcombine routine
with the ``ccdclip'' algorithm. As the images were not yet rotated to
align them on the sky, this clips out real objects as if they were cosmic
rays. This trial PSF was then subtracted from each of the original images.
These subtracted images were each rotated about the position of
$\epsilon$~Eri to align the positions on the sky and then combined by
taking a straight average of the values at each pixel, but with locations
near the main diffraction spikes or the obstructed edges of the 50CCD
aperture masked out of the average. We found that masking out a rather
wide strip (about 3.4" in the diagonal direction) around the main
diffraction spikes gave the best results.

A number of faint objects are clearly visible in the field of view. A
mask was created for each unrotated image that identifies pixels that are
affected by real objects on the sky.A final PSF for
$\epsilon$~Eri was then made in the same way as the initial PSF
image, but with these sky objects masked out before taking the average.
The new PSF was then subtracted from the shifted crj images, and the
subtracted files were again rotated into alignment and averaged after
masking out the diffraction spikes and aperture edges.

To summarize, we masked out the sky objects when averaging the unrotated
and unsubtracted images to create the PSF, and then masked out the
diffraction spikes and aperture edges when averaging the rotated and
PSF-subtracted images to create the image of the sky. In principle this
procedure could be iterated to refine the separation between
the sky and PSF images, but we will simply use the second version of the
sky and PSF images produced by this procedure.

\begin{figure}[htb]
\includegraphics[angle=0,scale=0.85]{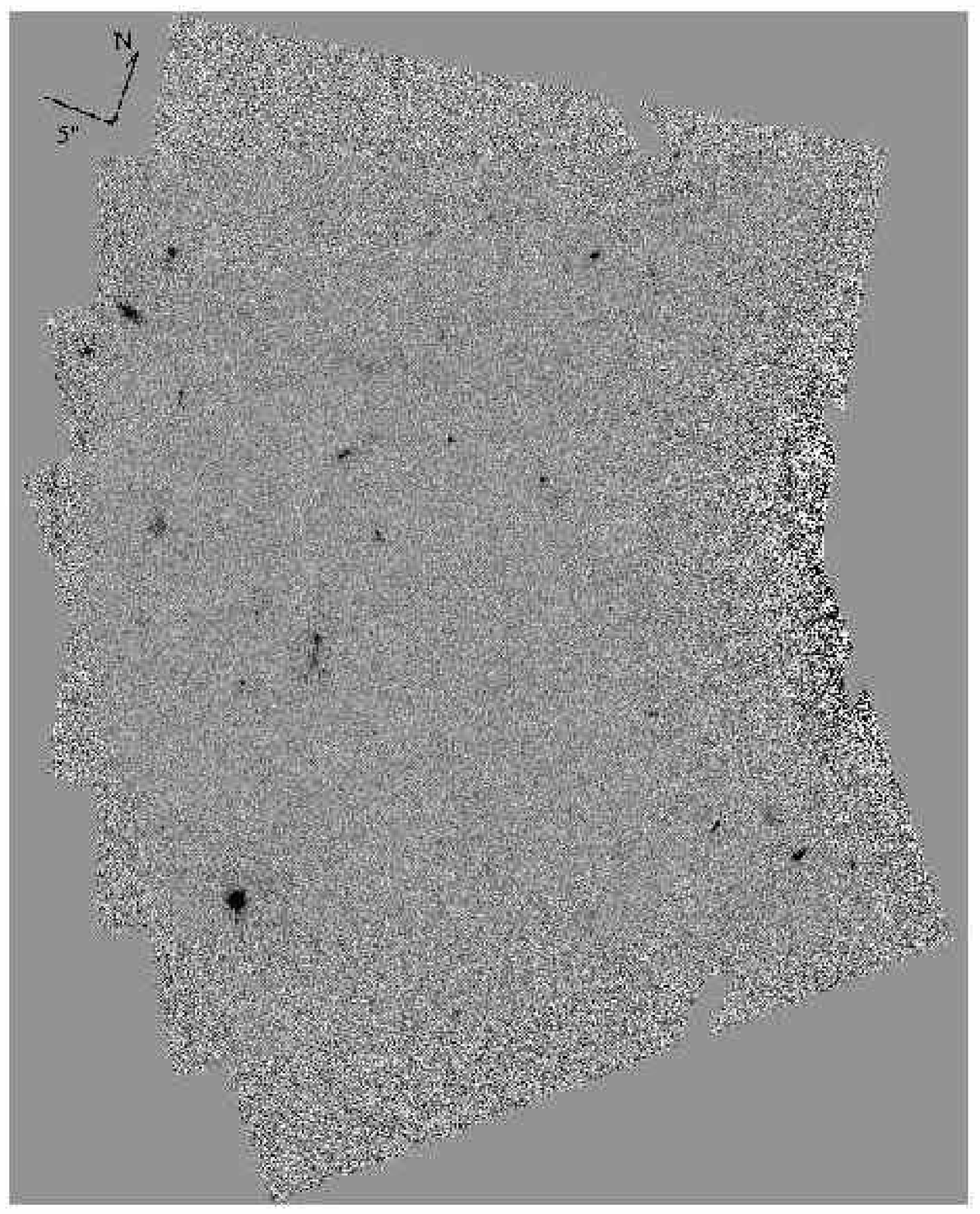}
\caption{Full mosaiced, roll subtracted image of the targeted field near
$\epsilon$~Eri. This, and other images of this field are aligned with the STIS
observation o6eo03020, with the +y direction aligned 20 degrees east of north.
The compass drawn in the upper left corner of this figure is 5" on a side.
\label{rollsub} }
\end{figure}

\placefigure{rollsub}

  \subsubsection{Direct Subtraction of $\epsilon$ and $\delta$ Eri's PSFs}

    \subsubsubsection{$\delta$ Eri PSF} The procedure used to produce the
$\delta$~Eri PSF was similar to that used to produce the $\epsilon$~Eri PSF. 
It was necessary to mask out a generous region around the brightest background
star in each of the two $\delta$~Eri images to avoid introducing obvious
artifacts in the subtraction.

    \subsubsubsection{Relative Normalization of the Two Stars}

Before subtracting the $\delta$~Eri PSF from the observations of
$\epsilon$~Eri, it is necessary to know both $r$, the relative
normalization of the two PSFs, as well as $b$, the sky background level.
The PSF subtracted images will then be calculated as $ I = (I_\epsilon -
b) - r(P_\delta-b) $.

To calculate the relative normalization we need to consider the spectral
energy distributions (SEDs) of the two stars. We will approximate these SEDs
by using the broad-band photometry (Table \ref{phottab}) taken from the
Lausanne online database (Mermilliod, Mermilliod, \& Hauck 1997). Longward of
the $V$ band, the two stars have very similar spectral distributions, but
$\epsilon$~Eri is a bit bluer at shorter wavelengths. We supplement this
photometry with IUE data for shorter wavelengths to provide a rough spectral
energy distribution for each star (Table \ref{fluxtab2}), which is then used
with SYNPHOT to predict STIS CCD imaging magnitudes in the 50CCD and F25ND3
filters. The predicted F25ND3 magnitudes are about 8\% brighter than
observed, but the predicted ratio of the two stars' F25ND3 count rates
matches the observed ratio to within 0.2\%. This gives us a fair measure of
confidence that the predicted flux ratio for STIS CCD imaging with the
unfiltered 50CCD aperture will also be correct.

\placetable{phottab}

\placetable{fluxtab2}

\placetable{rattable}

We have also constructed Tiny Tim (Krist 1993; 1995) models using the above
spectral energy distributions as input. Tiny Tim only calculates on-axis PSFs for
STIS, and does not calculate the PSF beyond ~4.5" radius. The ratio of the Tiny
Tim PSFs does show that, when normalized to the same total flux, the slightly
bluer $\epsilon$~Eri SED results in a PSF at 4.5" that is about 0.4\% lower than
that of $\delta$~Eri. Taking the predicted 50CCD ratio of 0.837, and then
assuming  the same difference between the predicted and observed ratios as was
found for the direct F25ND3 images ($0.837-0.845$), would give an expected ratio
of 0.839.  Extrapolating the radial color difference found in the Tiny Tim models,
we estimate an additional correction of about $-0.005$ near 17", yields an
estimated normalization ratio of 0.834. If this ratio is adopted, it is also
necessary to assume a mean sky background of $\approx$0.077 e$^-$/pixel/s if the
farthest parts of the two PSFs are to match. This is comparable to the expected
sky brightness.

\begin{figure}[htb]
\includegraphics[angle=0,scale=0.85]{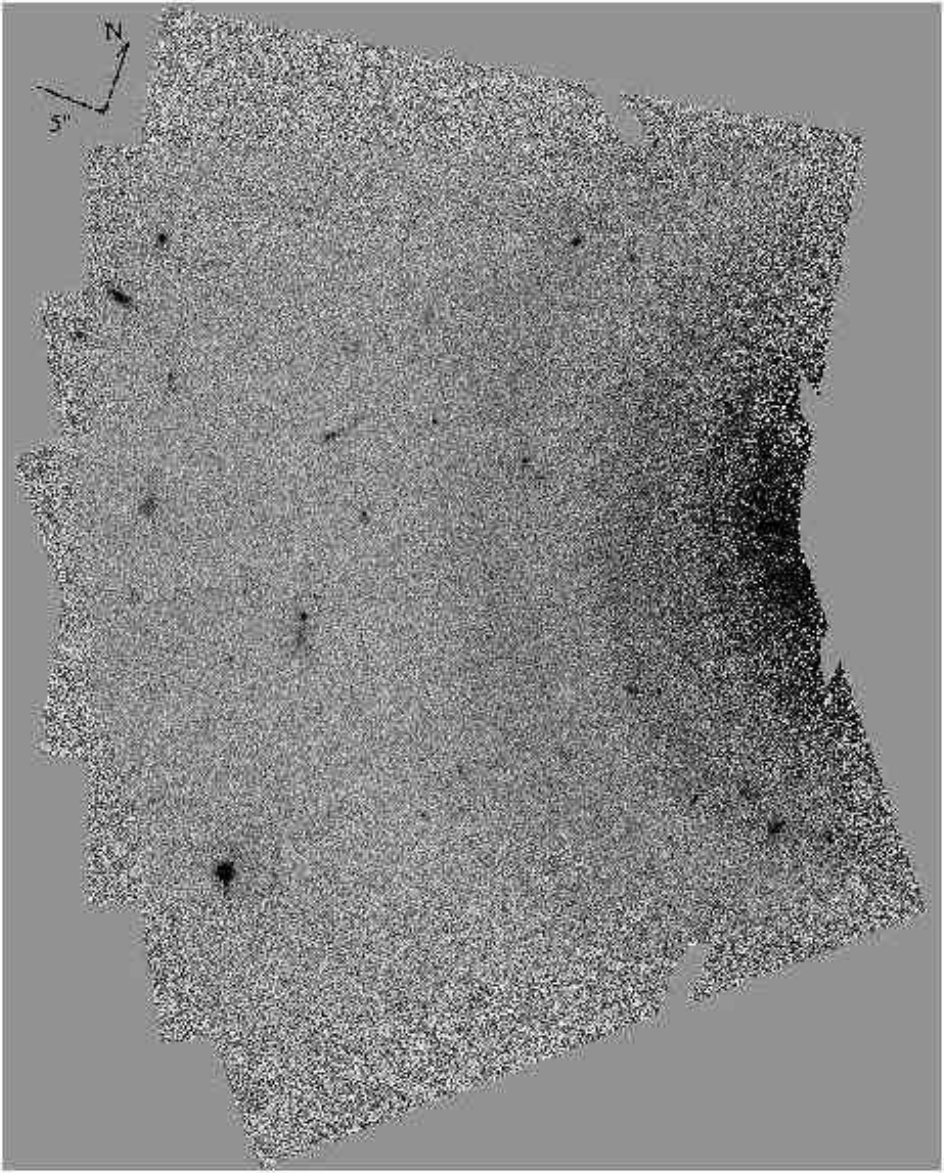}
\caption{Result of subtracting the two stars' PSF assuming a
normalization factor of 0.834. This normalization leaves a central
excess as well as two 5" wide rings with radii of approximately 18" and
28".
\label{delsub834} }
\end{figure}

\begin{figure}[htb] 
\includegraphics[angle=0,scale=0.85]{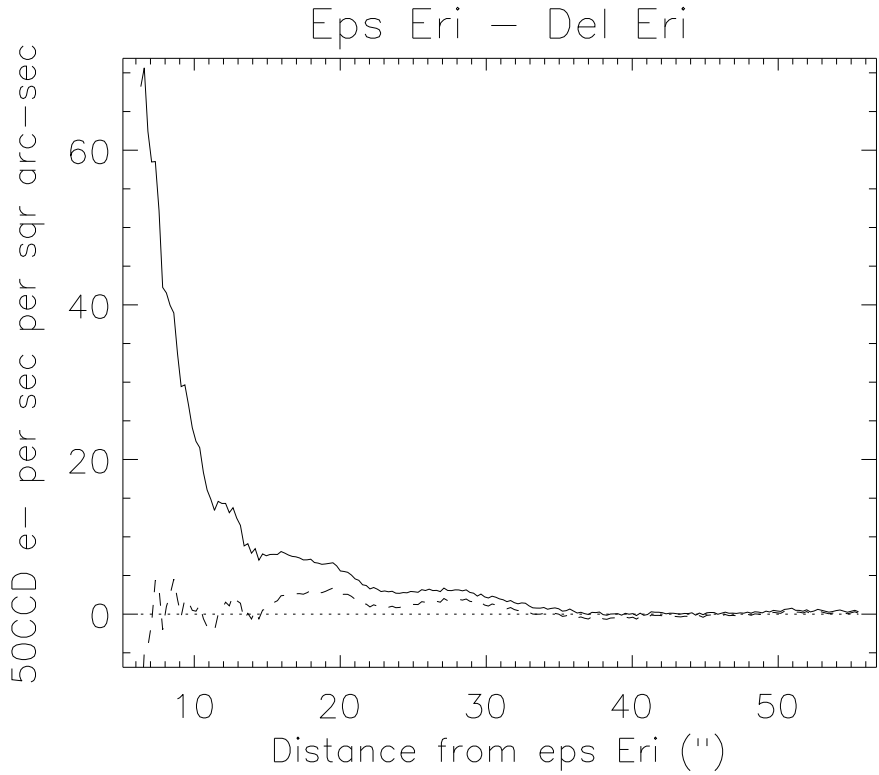}
\caption{The solid line shows the mean count rate (e-/arcsec$^{-2}$/s)
as a function of radius from $\epsilon$~Eri, after subtracting the
$\delta$~Eri PSF with a normalization factor of 0.834 (see
Fig.\ \ref{delsub834}). If we change this normalization factor to 0.842,
(dashed line) then the overall gradient is minimized, but the rings
remain, as shown in Fig.\ \ref{delsub842} \label{halo_cnts} }
\end{figure}

\begin{figure}[htb]
\includegraphics[angle=0,scale=0.85]{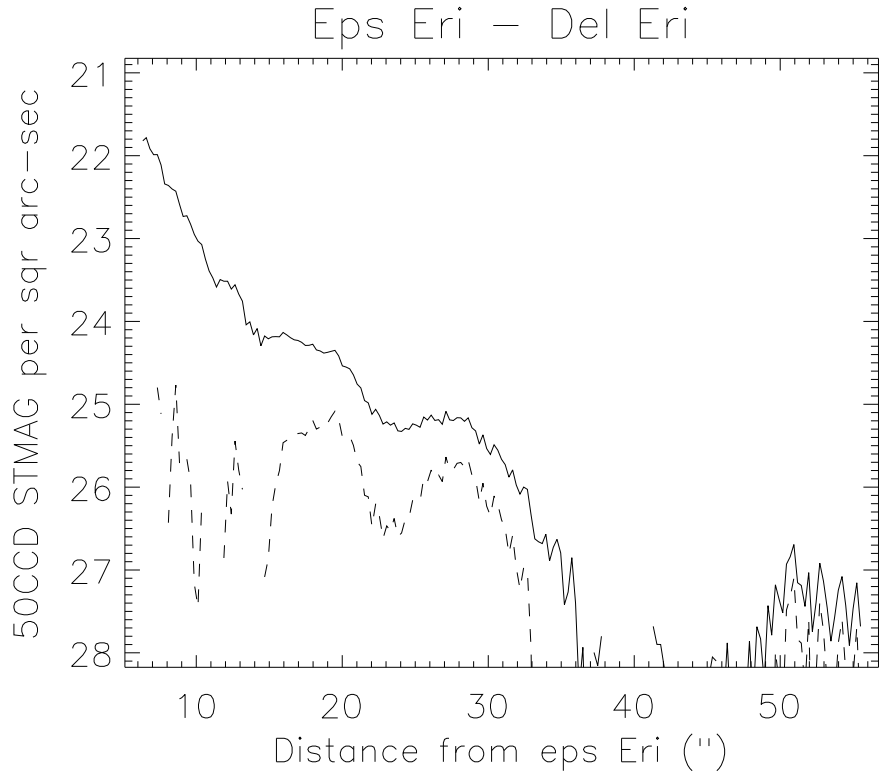}
\caption{The same as figure \ref{halo_cnts}, but in units of STMAGs per
arcsec$^{-2}$.
\label{halo_mags} }
\end{figure}

\begin{figure}[htb]
\includegraphics[angle=0,scale=0.85]{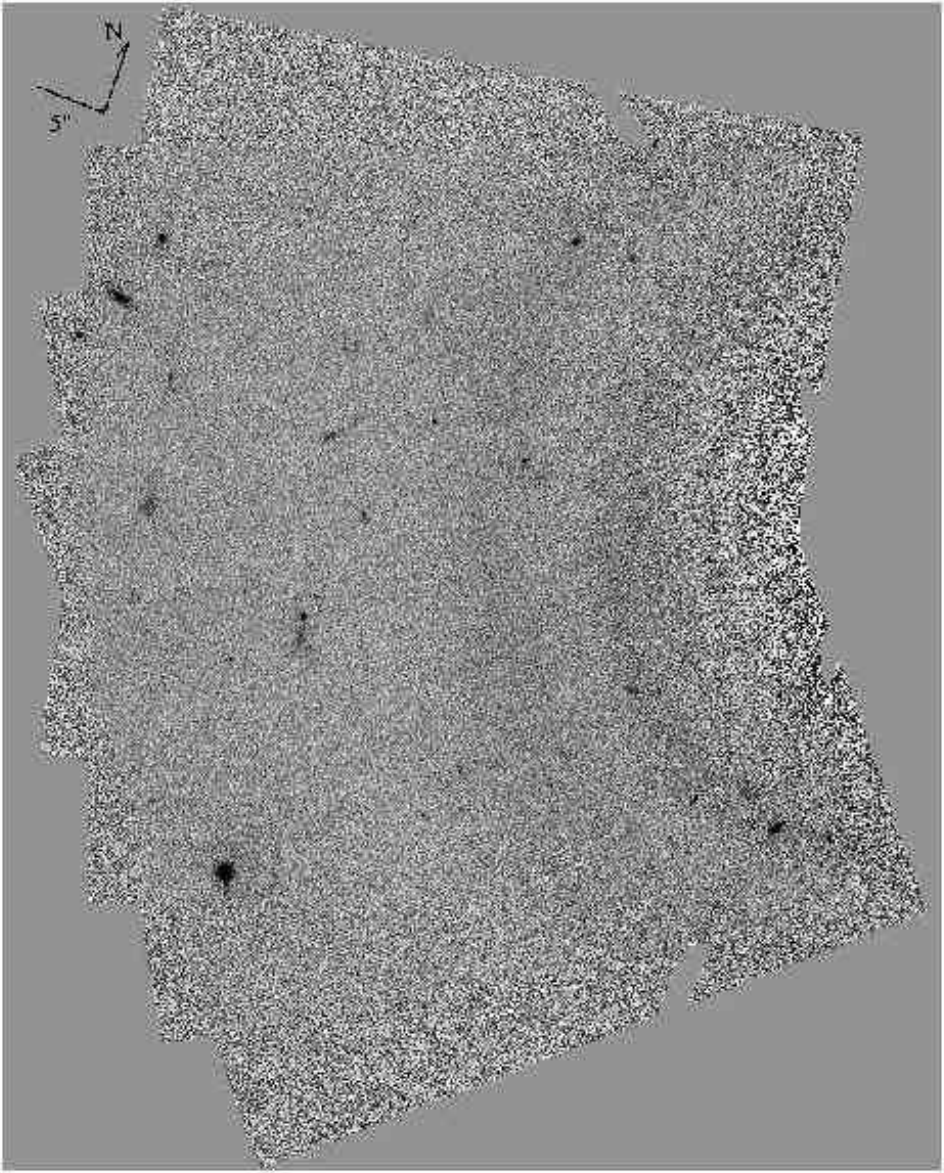}
\caption{Result of subtracting the two stars' PSF assuming a
normalization factor of 0.842. This normalization minimizes the mismatch
in flux close to the star, but the two broad rings remain.
\label{delsub842} } \end{figure}

The results of this subtraction are shown in Figure \ref{delsub834}. There
appears to be substantial excess light near $\epsilon$~Eri, amounting to
about 1\% of the unsubtracted PSF at the same radius, and a number of
concentric rings are visible. To illustrate this diffuse structure we took
the difference between the $\epsilon$ and $\delta$ Eri PSFs (with the sky
objects removed), and measure the mean surface brightness as a function of
radius (Figures \ref{halo_cnts} and \ref{halo_mags}).  We can minimize the
central halo around $\epsilon$~Eri by increasing the normalization constant
to about 0.842 (see Fig.\ \ref{delsub842}), but the concentric rings remain,
with a mean surface brightness of 25.5 to 25 STMAG/arcsec$^{-2}$. If we
assume the rings extend all the way around the star, this flatter
normalization corresponds to a total integrated STMAG of 17.4 for the rings.

\placefigure{delsub834}

\placefigure{halo_cnts}

\placefigure{halo_mags}

\placefigure{delsub842}

This structure seems suspiciously symmetrical to be consistent with the
observed clumpiness in the sub-mm observations, but the radial distribution
of the flux, with a broad hump between 15 and 30", is roughly consistent
with the radial dependence of the 850$\,\mu$m flux as shown in Figure 2 of
Greaves et al. The rings are broader than PSF subtraction artifacts
previously seen in STIS coronagraphic imaging (Grady et al.\ 2003), however
few images expose the far PSF wings this deeply. One example of ACS
coronagraphic observations of Arcturus did show that breathing induced
changes between visits can lead to a ring in the subtracted image with a
radius of 13" and a surface brightness 20.5 mag arcsec$^{-2}$
fainter than the star (see figure 5.12 of Pavlovsky et al.\ 2003). It may be
that the rings in our STIS data are a similar artifact. However, we will not
completely ignore the possibility that the rings might be real, and will
consider whether they could be plausibly modeled by dust in the
$\epsilon$~Eri system.

\section{Results}

\subsection{Comparison of the STIS images with the sub-mm
emission}

\subsubsection{Comparison with roll-subtracted images.}

\begin{figure}[htb]
\clearpage

\includegraphics[angle=0,scale=0.85]{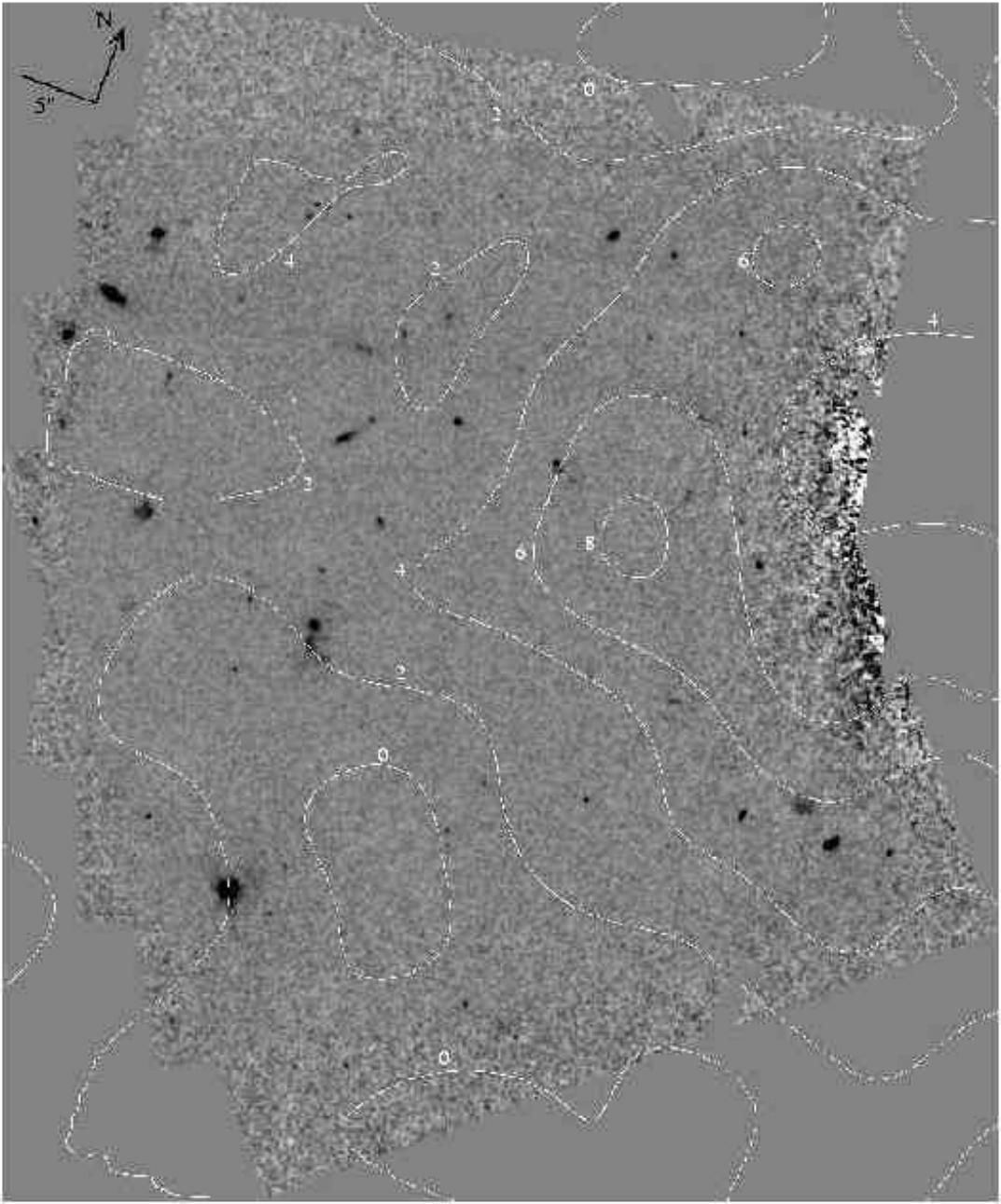}
\caption{A 5x5 boxcar smoothed version of the roll subtracted data is
compared with the sub-mm contour map, assuming the sub-mm emission
shares $\epsilon$~Eri's proper motion. The white numerals give
the value of Greaves et al.'s (1998) 850$\,\mu$m contour levels in units
of mJy per SCUBA beam area.\label{conmap} } \end{figure}

Figure \ref{conmap} shows the roll subtracted data from part of figure
\ref{rollsub}, but at two times the scale of that figure and after smoothing
with a 5x5 boxcar filter to suppress the small scale noise. Overlaid on
this figure is the sub-mm contour map of Greaves et al.\ (1998), positioned
assuming that the sub-mm emission shares the proper motion of
$\epsilon$~Eri. The inner contour level of the brightest sub-mm peak is
about 5" in diameter. 

\placefigure{conmap}

There are a substantial number of faint objects detected in the region of
the sub-mm emission, but the density of such objects is not appreciably
greater than elsewhere in the image. There is no apparent correlation
between these objects and the sub-mm flux, and most or all of these
objects are probably background galaxies or stars. Note that the orbital
period for material in the dust ring is about 500 years, and
the expected orbital motion is only about 0.2" per year.

\begin{figure}[htb]
\includegraphics[angle=0,scale=0.85]{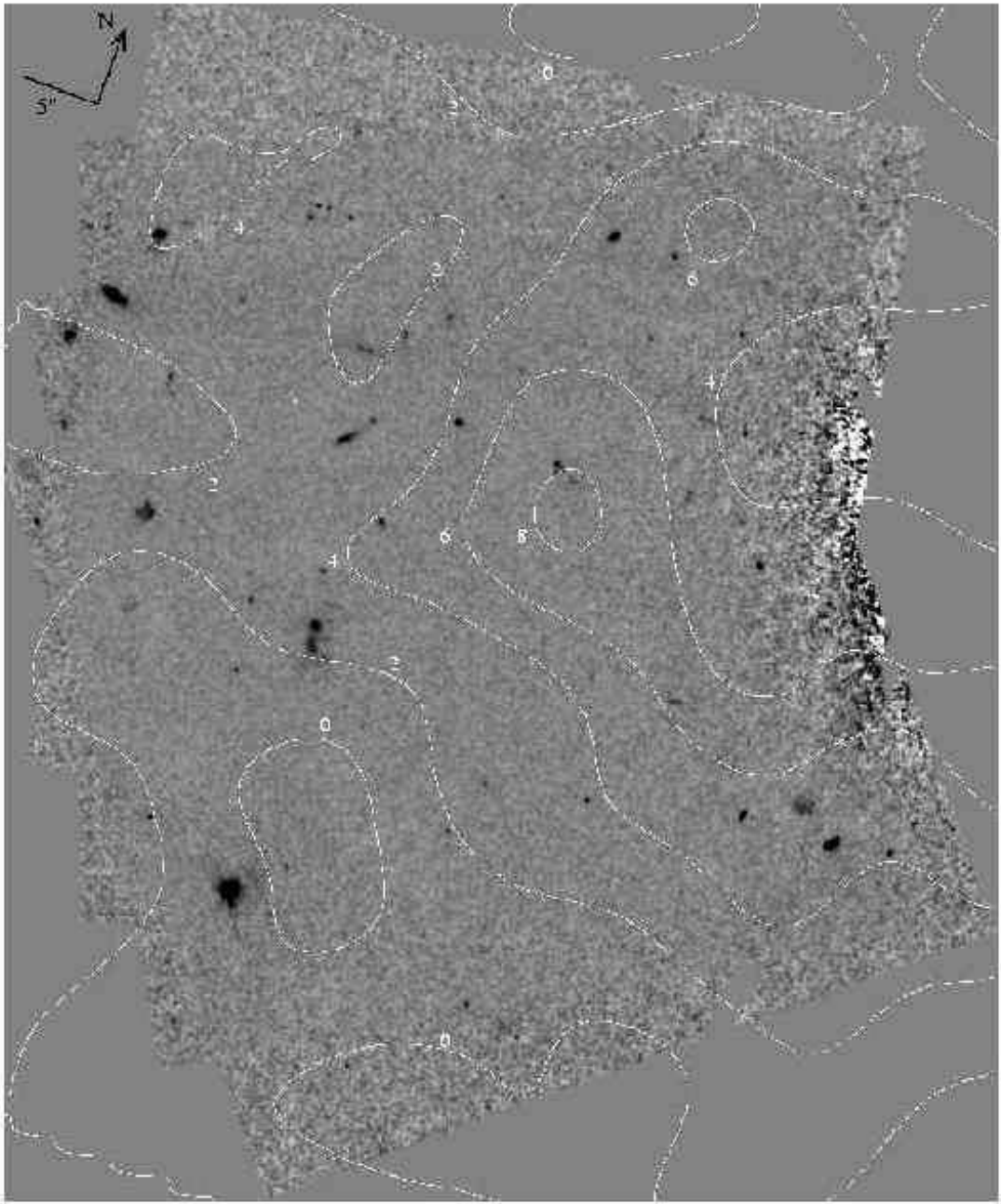}
\caption{The same as figure \ref{conmap}, but without correcting the
sub-mm contours for the proper motion of $\epsilon$~Eri. \label{conmap98}
} \end{figure}

Figure \ref{conmap98} shows the same data, but leaves the sub-mm contours
at where they were actually observed in 1997-1998; i.e., they are not
corrected for the $\approx$4" that $\epsilon$~Eri moved during the
intervening years. In this figure, the brightest sub-mm clump is just
outside the 2" sub-mm pointing uncertainty of what appears to be a
background galaxy with STMAG=24.7 (this brightness includes all the
clumps in this extended object). 

\placefigure{conmap98}

\begin{figure}[htb]
\includegraphics[angle=0,scale=0.85]{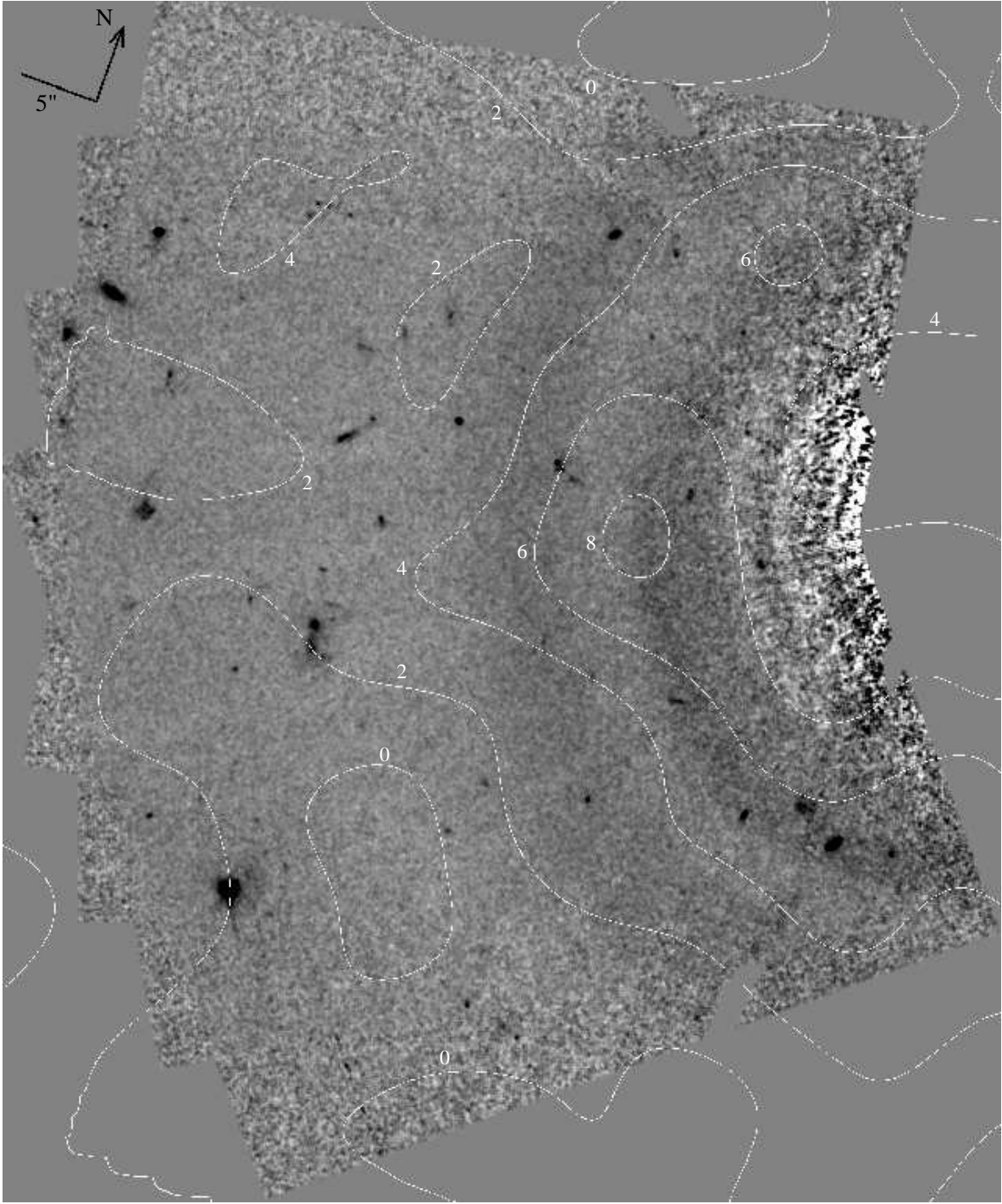}
\caption{The same as figure \ref{conmap}, except that this time the
contours are compared with the sky image made by subtracting the PSF
derived from the $\delta$~Eri observations (see Fig \ref{delsub842}). \label{conmapdz} } \end{figure}

Figure \ref{conmapdz} compares the proper motion corrected sub-mm
contours with a 5x5 boxcar smoothing of the image that results when
subtracting the $\delta$~Eri PSF from the $\epsilon$~Eri data. One of the
bright rings is roughly at the location of the sub-mm emission, but
there is no correlation with the bright knots.

\placefigure{conmapdz}

\subsection{Detected Objects Near $\epsilon$ Eri}

Nearly sixty distinct objects are visible in the field of view, including
about half a dozen within 20" of $\epsilon$~Eri. Most of these objects
appear to be slightly extended, and many have complex morphologies. There is
no apparent increase in their density at close distances to $\epsilon$~Eri,
and most are probably background galaxies. Two relatively bright point
sources are visible at distances of 51.5" (STMAG=19.6) and near 54"
(STMAG=22.2). A list of detected objects, their locations, and their
magnitudes is given in the appendix (Tab.\ \ref{objecttab}). For comparison,
we estimate, from the magnitudes and colors given by Chabrier et al.\ (2000)
for models of 1 Gyr old brown dwarfs with dusty atmospheres, that a $0.05
M_\odot$ brown dwarf in the $\epsilon$~Eri system would have a STIS 50CCD
STMAG of about 19, and a $0.03 M_\odot$ brown dwarf an STMAG of about 27.5.

\subsection{Point Source and Surface Brightness Detection Limits}

We measured the pixel-to-pixel rms variation in apparently blank regions of
the final composite roll-subtracted image as a function of radius from epsilon
Eri. The point-to-point rms noise varies  smoothly and can be fit as 
$\sigma\approx31/r^{3.16}+0.011$ e-/sec, where r is the distance in arc-sec
from epsilon Eri.  The measured noise is close to that predicted by a simple
noise model based on the read noise and the total number of counts in the PSF
as a function of radius. Within 30" of the star, the noise in the PSF
subtracted image is dominated by the Poisson noise of the star's PSF wings
(although the PSF has been subtracted, its Poisson noise still affects the
subtracted image). Since surface brightness of the stellar PSF increases
steeply closer to the star, our limiting magnitudes become significantly
brighter. At distances greater than 30", the accumulated read noise dominates
the total noise. The measured and predicted noise are listed in Table
\ref{noisetab}. This noise will be somewhat higher in regions which were not
observed at all four orients. Note that a total count rate of 1 e$^{-}$/s
corresponds to a 50CCD STMAG of 26.405, and that the magnitude of
$\epsilon$~Eri in these same units is 4.02.

For extremely red objects, as little as 10\% of the total point source counts
will fall into the central pixel of 50CCD images. For an unambiguous point
source detection, this central pixel should be at least 5$\sigma$ above the
rms noise. These assumptions lead to the detectability limits given in the final
column of Table \ref{noisetab}.

\placetable{noisetab}

For extended objects, the entire flux should be considered, but a 5 sigma
detection will still be required. If noise in different pixels was
uncorrelated, then the upper limit for detecting a fixed value of the surface
brightness would decrease as the square root of the area. For very large
areas, however, any low-spatial frequency noise sources would limit the 
practically achievable faint limit.

We empirically tested the real faint limit for extended sources by masking out
discrete sky objects and then comparing the measured background counts in a
number of separate  regions at various distances from $\epsilon$~Eri in the
roll subtracted image. Taking the standard deviation of the mean fluxes in
boxes of a given size at a given distance as the $1\,$sigma error in the
background measurement together with the Poisson noise from the potential
source, we derive the $5\,\sigma$ detection limits given in Table
\ref{noisetabe}. For boxes $<0.5$"$\times0.5$", we found about the variance
expected from scaling the point-to-point fluctuations by $1/\sqrt{n_{\rm
pix}}$, but for larger areas, the fluctuations were bigger than would be
expected from Poisson statistics. For example, when averaging over a
$3.5"\times3.5"$ box at 20", the measured pixel-to-pixel variance would imply
a $5\,\sigma$ detection limit of about 27.2 STMAG/arcsec$^{-2}$ -- about 1 mag
fainter than the directly measured limit. This difference could be due in part
to real sources on the sky, but the lack of a clear correlation of these
fluctuations with the sub-mm map or distance from the star, requires us to
treat such fluctuations as noise.

\placetable{noisetabe}

Note that at 20" from the  star, the shift in position from a 10 degree roll
change is about 3.5". Circular or arc-like structures much larger than the
roll separation will tend to be washed out in the roll-subtracted image.
Unfortunately, long narrow arc-like structures oriented in the tangential
direction are precisely the kind of structures that are most likely in
circumstellar debris disks. This limits the utility of the roll subtracted
image for detecting circumstellar structures much larger than a few
arc-seconds in extent.

\section{Modeling the dust in the $\epsilon$~Eri system}

\subsection{The size distribution of dust in circumstellar debris disks.}

Collisional fragmentation in a sufficiently dense debris disk will lead to
a collisional cascade that generates a wide spectrum of particle sizes.
Wyatt et al.\ (1999), discussed extensively the processes that influence
the resulting size distribution. Theoretical arguments predict that, when
collisions produce similar fragmentation at all size scales, the cascade
leads to a distribution of particle sizes $dn(a)\propto a^nda$ (Dohnanyi
1969; Tanaka, Inaba, \& Nakazawa 1996), with $n\simeq-3.5$. Such a
distribution has most of the mass concentrated in the biggest objects, but
with most of the surface area being dominated by particles near the lower
end of the size distribution. When the emissivity is constant for all
sizes, then the observable characteristics will then be dominated by the
smallest particles.

The distribution may, however, be considerably flattened if the orbital
evolution of the smaller particles affects them rapidly enough. The lower
cutoff to the distribution is set either by the size at which the time scale
for Poynting-Robertson driven orbital decay is smaller than the lifetime of
these particles against collisional creation, or when the smallest grains
can be blown out of the system by radiation pressure.

A small particle broken off a large object in a circular orbit will be
on an unbound hyperbolic orbit when the ratio of radiation to
gravitational forces $\beta_{\rm rad} > 0.5$. We can calculate
$\beta_{\rm rad}$ using the same Mie calculations and stellar spectrum
used to calculate the dust spectrum (see \S~\ref{dustcalc}). 

The time for Poynting-Robertson orbital decay is proportional to the ratio of
gravitation to radiative forces on a dust grain. The time
scale for Poynting-Robertson drag to change a circular orbit from  an radius
$r_1$ to $r_2$ is (Burns, Lamy, \& Soter 1979)
\begin{equation} t_{pr}=400(M_\odot/M_*)(r_1^2-r_2^2)/\beta_{\rm rad}\rm{\
yr}, \end{equation} 
where $r_1$ and $r_2$ are given in AU. For numerical comparisons we will
define  the time needed for the orbit to decay by 20\% of its initial
radius as the time scale for Poynting-Robertson drag; this is 36\% of the
time to decay to zero radius. At a distance of 60 AU from
$\epsilon$~Eri, and assuming the age of the system to be about 0.5 -- 1 Gyr, the
Poynting-Robertson effect would be expected to remove any grains
smaller than $\approx$100$\,\mu$m size unless the disk is dense enough
that the small particles are still being generated in collisions.

For the smallest particles in a debris disk, Wyatt et al.\ (1999)
estimated that the collisional time scale is of order
$t_{\rm{orb}}/4\pi\tau$, where $t_{\rm{orb}}$ is the orbital period and
$\tau$ is the effective face on optical depth (i.e., the geometric
filling factor). They also estimated that particles will only be
destroyed in collisions with particles larger than a factor of
$0.03\left[(M_\odot/M_*)(r/r_\oplus)\right]^{1/3}$  times their own size
(about 12\% at 60 AU from $\epsilon$~Eri), so we will define $\tau_d$ to
be a function of the particle size by integrating only over the
appropriate size range, and the time scale in years for fragmenting
collisions is then:
\begin{equation}
t_d={\sqrt{(r/r_\oplus)^3(M_\odot/M_*)}\over{4\pi\tau_d}} \simeq
 3.7\times10^5 \left[ \left({r\over 60 r_\oplus}\right)^{3/2} \left({M_*\over M_\odot}\right)^{-1/2} 
 \left({\tau_d \over 10^{-4}}\right)^{-1}\right]\,yr.
\end{equation}

For a dust model to be self consistent, this time scale should be
shorter than $t_{pr}$ for even the smallest particles in the model. 
We will see below that this condition is easily satisfied for the debris disk
around $\epsilon$~Eri.

\subsection{Calculation of dust models.}

\subsubsection{Calculating the spectrum of optically thin circumstellar
dust \label{dustcalc}}

Given an assumed grain composition and the associated wavelength dependent
optical constants, we performed standard Mie theory calculations for
spherical particles covering a wide range of radii $a$ using the code of
Wiscombe (1979; 1980). This gives, among other quantities, the standard
scattering and absorption coefficients $Q_{\lambda,sca}$ and
$Q_{\lambda,abs}$. (The emission coefficient at each wavelength
$Q_{\lambda,em} =Q_{\lambda,abs}).$

For the dust calculations we adopt the flux from a standard solar
abundance Kurucz model atmosphere with $T_{\rm eff} =5180K$ and $\log
g$=4.75, normalized to a total stellar luminosity of  0.35$\,L_\odot$.
The dust is assumed to be optically thin, and the temperature of each
dust grain can be determined by the equilibrium between absorbed stellar
radiation and thermal emission. Once the temperature of the grain is
determined, then we can calculate the total light from
a dust grain at each wavelength as the sum of the scattered starlight
and the thermal emission. We will assume that the dust grain is small
enough that the thermal emission is isotropic. The scattered light,
however, will be highly anisotropic. The angular phase function
$f(\theta)$ for this scattering can be easily derived from the Mie
calculation, and we normalize this function so that $f(\theta)=1$ for
isotropic scattering. Then, when viewed at a scattering angle $\theta$,
the apparent flux of the dust grain will be

\begin{equation}F_{\lambda,g}=\pi a^2 Q_{\lambda,sca}
F_{\lambda,d}f(\theta) + 4\pi a^2 \pi B_\lambda(T)Q_{\lambda,em}\end{equation}

This result will be used below to calculate the expected optical spectrum for
a given dust distribution.

\subsubsection{Dust composition and porosity}

The observable dust particles in circumstellar debris disks are presumed
to be collisionally produced fragments of larger objects resembling
those in the solar system's Kuiper belt. These larger objects were
probably formed as very porous agglomerations of interstellar grains
early in the history of the system.  

For the composition and optical properties of these grains we will adopt the
model of interstellar and circumstellar dust developed by Li \& Greenberg
(1997; 1998), which assumes that the dust grains are a mixture of silicates,
organic refractories, and voids, with some fraction of the voids possibly
filled in by water ice. The Bruggemann mixing rule (Kr\"ugel 2003)  is used
to calculate the effective optical constants for the resulting mixtures from
the optical constants of organic refractories and amorphous silicates given
by Li \& Greenberg (1997), and that of vacuum.

Wyatt \& Dent (2002) were able to fit the sub-mm and IRAS data for the
debris ring around the A3V star Fomalhaut by assuming solid (non-porous)
grains consisting of 1/3 silicate and 2/3 organic refractory material by
volume, and a grain size distribution close to the theoretically
expected $n^{-3.5}$ that extended down to the radiation blowout limit
for that star.  While they could not completely exclude models with some
degree of porosity in the grains, they argued that the collisional
fragmentation should have resulted in significant compaction of the
grains despite the high porosity expected in the primordial parent
bodies. 

In contrast, Li \& Greenberg (1997; 1998), and Li \& Lunine (2003ab) favor
models for the debris disks around HD 141569A, $\beta$~Pictoris, and
HR~4796A that assume highly porous grains, with vacuum fractions, $P\approx
0.7$ to 0.9. In some cases these require a dust size distribution close to
$dn\propto a^{-3}da$, significantly flatter than the theoretically expected
$dn\propto a^{-3.5}da$.

Li, Lunine, \& Bendo (2003) have recently fit such a model to the
available data for $\epsilon$~Eri. They find that a model
assuming highly porous particles, and a rather flat size distribution,
can provide an excellent fit to both the IRAS and sub-mm data. They
assumed that the same dust-size distribution function applies at all
distances from the star, and only varied the total number density of
grains as a function of radius to match the distribution of the observed
850$\,\mu$m flux. It is not clear whether or not this is realistic.
Moro-Mart\'in's \& Malhotra's (2002; 2003) dynamical studies of dust
produced in our own Solar System's Kuiper belt found that the size
distribution function is expected to change substantially as a function
of radius. However, the Li et al.\ (2003) model does give an excellent
fit to both the total sub-mm and IRAS fluxes observed in the
$\epsilon$~Eri system.

This model assumes $dn \propto a^{-3.1}da$, lower and upper size limits to
the distribution ,$a_1=1\,\mu$m, and $a_2=1\,$cm, a porosity, $P=0.9$. The
solid portion  of the grains is assumed to consist of an organics/silicate
mix in a 58:42 ratio by volume. The radial distribution of dust density is
modeled as a Gaussian centered at 55 AU, with a FWHM of 30 AU. We repeated Li
et al's calculations for the flux from such a dust distribution, but also
included the contribution of the scattered light, assuming a scattering angle
of 90 degrees (i.e., a face on disk, as indicated by the morphology of the
850 $\mu$m flux). 

\begin{figure}[htb]
\includegraphics[angle=0,scale=0.85]{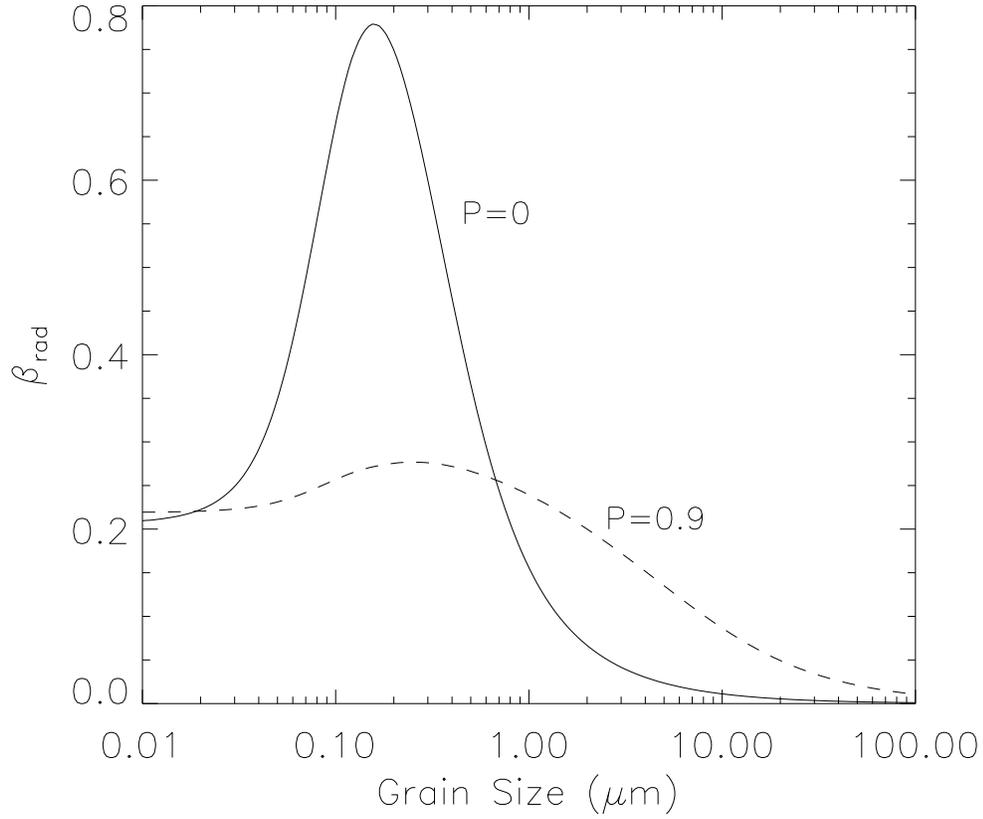}
\caption{The ratio of radiative to gravitational forces, $\beta_{\rm{rad}}$
in the $\epsilon$~Eri system for the porous grains assumed by Li et al.\
(2003) are plotted as a function of grain size (dashed line). For comparison,
$\beta_{\rm{rad}}$ for solid grains of the same composition is also plotted
(solid line). \label{bradplot}}
\end{figure}

\begin{figure}[htb] 
\includegraphics[angle=0,scale=0.85]{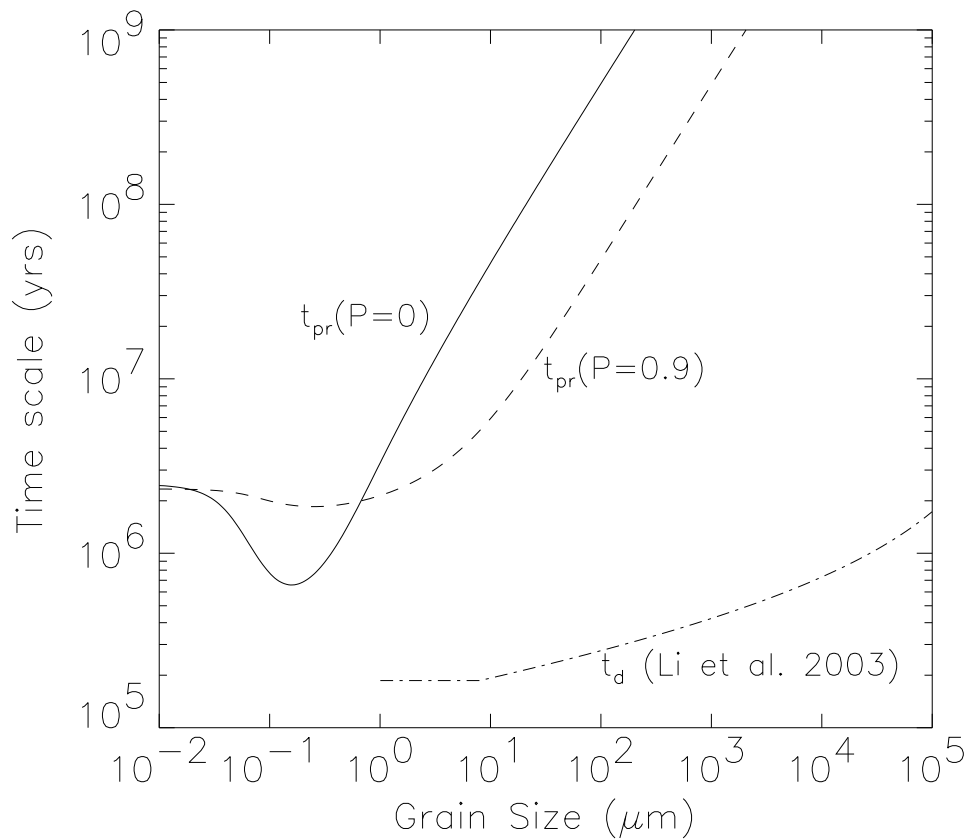}
\caption{The Poynting-Robertson orbital decay time-scale is compared for
solid ($P=0$) and porous ($P=0.9$) grains. This time scale is always much
larger than the time-scale for the fragmentation rate of grains ($t_{\rm{d}}$)
we calculate for the parameters of the Li et al.\ (2003) dust
model.\label{tplot}} 
\end{figure}

It is instructive to examine the radiative forces and time scales for this
model of the dust distribution. Figure \ref{bradplot} shows
$\beta_{\rm{rad}}$, the ratio of radiative to gravitational forces, for dust
grains of the modeled composition in the $\epsilon$~Eri system (see also
Figure~3 of Sheret et al 2003). At large
grain sizes, the low density of porous grains substantially increases the
effects of radiation pressure relative to that on solid grains of the same
composition and size. At small sizes, however, the porous grains no longer
effectively scatter radiation, and $\beta_{\rm{rad}}$ drops
below that of solid grains, never becoming large enough
($\beta_{\rm{rad}}>0.5$) for such grains to be efficiently blown out of the
system.  In reality, it is likely that the grains become less porous as they
are fragmented to very small sizes. In any case, the time scale for
fragmenting collisions expected for the model of Li et al (2003) is much
shorter than $t_{\rm{pr}}$ at all sizes (Fig.~\ref{tplot}), so small grains
should be abundant, but predicting the detailed distribution of grain size
and porosities at the lower end of the distribution will be difficult.

\placefigure{bradplot}

\placefigure{tplot}

Attempts made to fit solid grains models ($P=0$) have not resulted in as
good a fit as the Li et al.\ (2003) model. However, Sheret, Dent, \& Wyatt
(2003) found that a simple model of a thin ring at 60 AU with solid
silicate/organic grains, $a_1=1.75\,\mu$m, $a_2=5\,$m, and $dn \propto
a^{-3.5}da$, fits the IR and sub-mm data with a reduced $\chi^2$ of 2.8.

\subsubsection{Optical Brightness of dust models normalized to mean sub-mm
flux}

Near 55 AU from $\epsilon$~Eri the typical $850\,\mu$m surface brightness
observed by Greaves et al.\ is about 0.02 mJy/arcsec$^{-2}$. After subtracting
the mean emission of the ring,  the 850$\,\mu$m flux in the brightest clump
totals to about 2.6 mJy. If the angular extent of the clump is comparable to
the SCUBA beam-size, this amounts to a surface brightness enhancement of an
a additional 0.015 mJy/arcsec$^{-2}$.

\begin{figure}[htb]
\includegraphics[angle=0,scale=0.85]{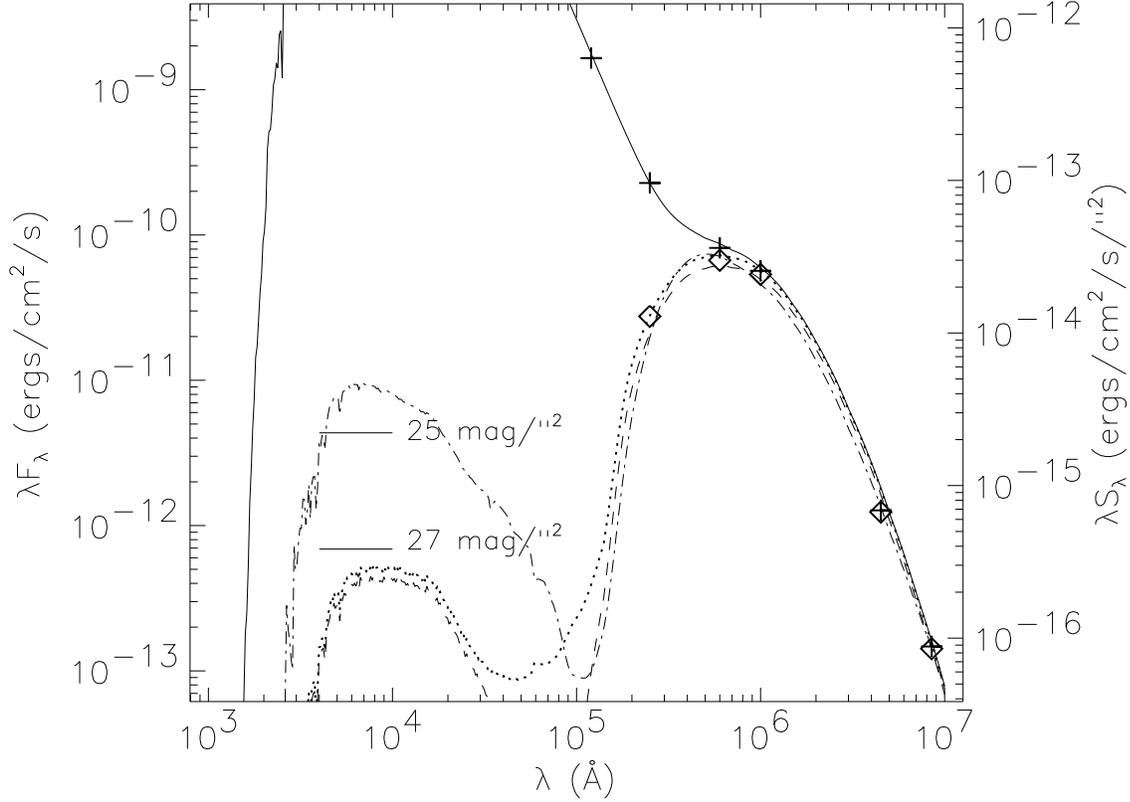}
\caption{The dotted line shows our calculation of the total spectral
energy distribution for Li et al's (2003) favored model for the dust ring
around $\epsilon$~Eri, including the contribution from scattered optical
light. The solid line shows this model added to the stellar SED. The (+) show
the observed IRAS and sub-mm flux observations, and the (diamonds) show the
flux values after subtracting the star light.  The dashed line, shows the SED
of the surface brightness of this model at 55 au (surface brightness values
given on right axis), with the scales shifted to overlap the curves at
850$\,\mu$m. The dash-dot line shows the SED for the parameters of the thin
ring solid grain model of Sheret et al.\ (2003). Also marked are the
approximate surface brightness levels corresponding to STIS 50CCD brightnesses
of 25 and 27 magnitudes/arcsec$^{-2}$.  These levels roughly correspond to the
brightness of the rings seen after the direct PSF subtraction, and the upper
limit to the surface brightness determined for the roll subtracted PSF image
respectively. \label{fig_model1} } 
\end{figure}

In Figure \ref{fig_model1}, we show, for a dust model with the parameters of
Li et al.\ (2003), the calculated total spectral energy distribution for the
whole dust cloud, and also for just the dust at 55 AU. Normalizing this
model to an 850$\,\mu$m surface brightness at 55 AU of
0.02$\,$mJy/arcsec$^{-2}$, yields a predicted STIS 50CCD surface brightness of
27.9 STMAG/arcsec$^{-2}$ (equivalent to $6.5\times10^{-4}$ cnts/pixel/sec). 

If we instead assume the parameters of the Sheret et al.\ (2003) solid grain
model, we predict an optical 50 CCD surface brigtness of 24.5 STMAG --
rather {\em brighter} than the rings in the directly subtracted image.

\placefigure{fig_model1}

In the roll subtracted image, any tangential feature large than the roll
change between images will show up in the PSF rather than in the sky image,
and so circular rings would be invisible. The bright clump might be detectable
if it were not too diffuse or too spread out in the tangential direction.  If
the dust parameters in the clump are the same as in the rest of the ring, we
would predict a total 50CCD optical brightness of 22.6 STMAG.  Spread over a
SCUBA beam area, this would give a surface brightness of  about 28
STMAG/arcsec$^{-2}$ -- well below our most optimitic detection limits of
$\approx 27$ STMAG/arcsec$^{-2}$.  The clump would have had to be concentrated
within an area of no more than 1/4 of the SCUBA beam size before we would have
expected to see it.

This assumes that the dust distribution in the clump is the
same as that of the ring as a whole. If the clump was created by
resonant interactions of dust with a Neptune-like planet in the
$\epsilon$~Eri system, smaller particles, which are strongly perturbed
by radiation pressure, may be less likely to collect in the same
resonances. If, for the enhancement in the clump, we change the lower
limit of the size distribution to be $150\,\mu$m, and again normalize
to 0.015 mJy/arcsec$^{-2}$ at $850\,\mu$m, then the predicted surface
brightness of the clump drops to about 29.8 STMAG/arcsec$^{-2}$
($1.5\time10^{-4}\,$cnts/pixel/s). Such a model also substantially
reduces the clump's contrast against the rest of the ring in the
thermal IR.

The rings seen in the directly subtracted image ($\epsilon$~Eri$ -
\delta$~Eri), with a surface brightness of about 25 STMAG/arcsec$^{-2}$, are
much brighter than the predictions of Li et al.\ (2003), but slightly fainter
than predicted by the model of Sheret et al (2003). If the rings are PSF subtraction
artifacts, they are bright enough to obscure the expected signal from the
dust. If real, they would imply a much larger abundance of small grains that
scatter efficiently in the optical than does the model of Li et al.
The lack of obvious counterparts in the optical ring corresponding to the
sub-mm clumpiness might be explained by the very different dynamical
behavior of the small grains. 

Unfortunately available modeling of the HST/STIS PSF at distances of 20" is
inadequate to provide a clear answer regarding the reality of the ring-like
features seen after the subtraction of the two stars' PSFs. 

\subsubsection{Variations of models}

Both the models of Li et al.\ (2003) and Sheret et al.\ (2003) assume that
all excess IR and sub-mm flux in the $\epsilon$~Eri system is due to a
single dust distribution that is well traced by the 850$\,\mu$m emission. 
If a substantial fraction of the IR-excess is instead due to an inner
zodiacal cloud of particles that contributes little at 850$\,\mu$m, then
constraints on possible models are considerably relaxed, although the IRAS
measurements will still provide an upper limit to the allowed IR flux from
the 55 AU ring. 

For example, if we change the model of Li et al.\ (2003) by simply
allowing the upper limit of the size distribution to extend to
$10^7\,\mu$m instead of $10^4\,\mu$m, and adjust the overall
normalization to again match the observed $850\,\mu$m flux, then both
the predicted IRAS band and optical fluxes
drop by a factor of about 4. If we instead assume Li et al's
parameters, but with solid rather than porous grains, the predicted
IRAS fluxes drop by a factor of two to three, but the optical surface
brightness {\em increases} by a factor of four. However, without better
information on the spatial distribution of the 10 to $100\,\mu$m flux
it is difficult to choose among the different possible models. Such
information will eventually be provided by  SIRTF images of the
$\epsilon$~Eri system. However, there is still unique information about
the distribution of small grains that would be provided by direct
detection of the optical scattered light that cannot be obtained from
even the most detailed observations of the thermal dust emission.

The simple models discussed here clearly have some limitations. Real dust does
not consist of the perfectly smooth spheres assumed in Mie theory, but will
have considerable surface roughness. For example, Lisse et al.\ (1998) in a
study of cometary dust found that considering the expected fractal structure
of the dust could increase the optical scattering at 90 degrees by as much as
a factor of three. Also, the porosity is unlikely to be the same for all grain
sizes. Even if the parent bodies are highly porous, at sufficiently small
sizes the dust may either be significantly compacted by collisions, or will
have broken up into smaller but more solid component particles.  However,
currently there are insufficient observational data for the $\epsilon$~Eri
dust ring to constrain the additional free parameters needed by such models.
This situation will improve substantially when {\em Spitzer Infrared
Telescope} images of this system become available. 

\section{Conclusions}

Our deep optical observations of the $\epsilon$~Eri sub-mm ring have not
provided clear evidence for detection of an optical counterpart. The upper
limits measured are consistent with existing models of the dust in the sub-mm
ring, and provide some constraints on the nature and amount of the smallest
dust grains. Our optical limits should provide tighter constraints once {\em
Spitzer Infrared Telescope} images are available for this system.

We found approximately 59 objects between 12.5" and 58" from $\epsilon$~Eri,
with brightnesses between 19.8 and 28 magnitude (STIS/50CCD STMAG). If any
of the more compact of these objects were  associated with the $\epsilon$
Eri system, they would correspond to brown dwarfs of $\approx$ 0.03 to 0.05
$M_\odot$.  However, it is much more likely that the majority of these
objects are background stars and galaxies unrelated to the $\epsilon$~Eri
system. A second epoch HST observation of comparable depth would immediately
determine whether any of these objects shares  $\epsilon$~Eri's 0.98 "/year
proper motion.

\acknowledgments

Support for proposal GO-09037 was provided by NASA through 
grants from the Space Telescope Science Institute, which is operated
by the Association of Universities for Research in Astronomy, Inc.,
under NASA contract NAS 5-26555.

\appendix

\section{Appendix}

Table \ref{objecttab} contains a list of detected objects. Some of the listed
objects close into the star may well be noise or PSF subtraction artifacts,
while other faint but real objects may still be omitted from this list. For
each object the J2000 coordinates at epoch 2002.071, the distance from
$\epsilon$~Eri, the approximate size of the object, and the total brightness
in STIS 50CCD STMAG units are listed.  For compact objects, the size given is
the FWHM from a Moffat fit, while for more extended objects the dimensions
given are an approximate estimate. One STIS CCD pixel corresponds to about
0.05071" on the sky. 

Macintosh et al.\ (2003), performed a $K$ band adaptive optics search for close
companions  around $\epsilon$~Eri, and found 10 candidates, although none are proper
motion companions to $\epsilon$~Eri.  Four of these objects lie in our field of view. 
Objects \#4 and \#6 correspond to extended galaxies and are noted in the table
below. Their objects \#5 and \#9 have no optical counterparts.

\placetable{objecttab}

\clearpage

\clearpage

\clearpage

 \clearpage

\clearpage

\clearpage

\clearpage

\clearpage

\clearpage

\clearpage

\clearpage


\begin{deluxetable}{ccll}
\tablecaption{IR and sub-mm measurements of dust around $\epsilon$~Eri.
\label{fluxtab}}
\tablehead{Wavelength & Dust Flux & Source & Comments\\  
           ($\mu$m)    &  (mJy)  & & \\}
\startdata
1200 & $21.4\pm5.1$ & Sch\"utz et al 2004&\\
850  & $40\pm3$   & Greaves et al.\ & \\
450  & $185\pm103$& Greaves et al.\ & \\
100  & $1780$     & {\em IRAS}    & Photospheric Flux Subtracted  \\
 60  & $1340$     & {\em IRAS}    & Photospheric Flux Subtracted  \\
 25  &  $270$     & {\em IRAS}    & Photospheric Flux Subtracted  \\
\enddata
\end{deluxetable}

\clearpage

\begin{deluxetable}{cccccc}
\tablecaption{Deep Offset 50CCD Observations\label{obstable}}
\tablehead{Target & dataset & RA & Dec & y-axis orient  & Expected clump\\
                  &         &    &     &  ($^\circ$e of N)  &location (pixels)    }
\startdata
 $\delta$ Eri OFF1  & o6eo01020 &03 43 16.95 &$-$09 45 52.09 & 10.0544 &\ldots  \\
 $\epsilon$ Eri OFF1& o6eo02020 &03 32 57.75 &$-$09 27 35.20 & 10.0546 & 735, 446 \\
 $\epsilon$ Eri OFF2& o6eo03020 &03 32 57.66 &$-$09 27 40.42 & 20.0547 & 729, 515 \\
 $\epsilon$ Eri OFF3& o6eo04020 &03 32 57.50 &$-$09 27 45.31 & 30.0548 & 734, 584 \\
 $\epsilon$ Eri OFF4& o6eo05020 &03 32 57.29 &$-$09 27 49.74 & 40.0550 & 752, 651 \\
 $\delta$ Eri OFF4  & o6eo06020 &03 43 16.49 &$-$09 46 06.64 & 40.0548 &\ldots   \\
\enddata
\end{deluxetable}
\clearpage

\begin{deluxetable}{lcc}
\tablecaption{F25ND3 Observations\label{nd3images}}
\tablehead{\colhead{Target} & \colhead{dataset}& \colhead{expo.\ time}\\
                  &        & \colhead{(s)}   }
\startdata
 $\delta$ Eri  & o6eo01010 & 0.4 \\
 $\epsilon$ Eri& o6eo02010 & 0.6 \\
 $\epsilon$ Eri& o6eo03010 & 0.6 \\
 $\epsilon$ Eri& o6eo04010 & 0.4 \\
 $\epsilon$ Eri& o6eo05010 & 0.4 \\
 $\delta$ Eri  & o6eo06010 & 0.6 \\
\enddata
\end{deluxetable}
\clearpage

\begin{deluxetable}{lcc}
\tablecaption{Photometry of $\epsilon$~Eri and $\delta$~Eri\label{phottab}}
\tablehead{band & $\epsilon$~Eri & $\delta$~Eri }
\startdata
 $V      $ & 3.726  & 3.527 \\
 $B-V    $ & 0.882  & 0.922 \\
 $U-B    $ & 0.584  & 0.686 \\
 $V-R_c  $ & 0.504  & 0.505 \\
 $R_c-I_c$ & 0.440  & 0.434 \\
 $V_c-I_c$ & 0.940  & 0.939 \\
 $J-K    $ & 0.55  & 0.56 \\
\enddata
\end{deluxetable}
\clearpage

\begin{deluxetable}{lccc}
\tablecaption{Adopted flux distributions for $\epsilon$~Eri and
$\delta$~Eri for PSF calculations.\label{fluxtab2}}
\tablehead{$\lambda$ (\AA) & $\epsilon$~Eri & $\delta$~Eri & ratio }
\startdata
2700     & $ 3.67\times10^{-12}$ & $2.75\times10^{-12}$ & 1.335\\
3000     & $ 1.15\times10^{-11}$ & $1.11\times10^{-11}$ & 1.036\\
3646.235 & $ 3.63\times10^{-11}$ & $3.82\times10^{-11}$ & 0.950\\
4433.491 & $ 8.74\times10^{-11}$ & $1.01\times10^{-10}$ & 0.865\\
5492.883 & $ 1.14\times10^{-10}$ & $1.37\times10^{-10}$ & 0.832\\
6526.661 & $ 1.08\times10^{-10}$ & $1.30\times10^{-10}$ & 0.831\\
7891.114 & $ 8.92\times10^{-11}$ & $1.07\times10^{-10}$ & 0.833\\
12347.43 & $ 4.02\times10^{-11}$ & $5.02\times10^{-11}$ & 0.801\\
22094.22 & $ 8.38\times10^{-12}$ & $1.06\times10^{-11}$ & 0.791\\
\enddata
\end{deluxetable}
\clearpage

\begin{deluxetable}{lccc}
\tablecaption{STIS CCD Imaging Photometry}
\tablehead{Filter & $\epsilon$~Eri & $\delta$~Eri & Flux ratio ($\epsilon/\delta$)\label{rattable}}
\startdata
50CCD (predicted)  &4.023 & 3.830 & 0.8371 \\
F25ND3 (predicted) &4.038 & 3.842 & 0.8345 \\
F25ND3 (observed)  &3.956 & 3.762 & 0.837 \\
\enddata
\end{deluxetable}
\clearpage

\begin{deluxetable}{cccc}
\tablecaption{50CCD Point Source Detection Limits in Roll Subtracted Image\label{noisetab}}
\tablehead{distance &  1$\sigma$ measured noise&  1$\sigma$ predicted noise& 5$\sigma$ point source \\
           (")      &  (e-/pixel/s)            &  (e-/pixel/s)            & limiting mag  }
\startdata
 7                  &   0.064 &   0.052     &   25.0       \\
 8                  &   0.055 &   0.043     &   25.3       \\  
 9                  &   0.041 &   0.037     &   25.5       \\  
10                  &   0.033 &   0.032     &   25.7       \\  
12                  &   0.023 &   0.026     &   26.1       \\  
15                  &   0.017 &   0.021     &   26.3       \\  
20                  &   0.013 &   0.017     &   26.6       \\  
30+                 &   0.012 &   0.015     &   26.7       \\  
\enddata
\end{deluxetable}
\clearpage

\begin{deluxetable}{cccccc}
\tablecaption{50CCD Extended Source 5$\,\sigma$ Surface Brightness Limits in Roll Subtracted Image\label{noisetabe}}
\tablehead{ distance from star& \multispan5{5$\,\sigma$ Limiting Surface Brightness (STMAG/arcsec$^{-2}$) vs.\ box size }\\
            (")&$0.5" \times 0.5" $  &  $1" \times 1"$   &  $3.5" \times 3.5"$ &  $5" \times 5"$ &}
\startdata
15   &  $ 24.85 $        &  $ 25.34 $    & $ 26.61 $       &  $ 27.29 $& \\
20   &  $ 25.06 $        &  $ 25.60 $    & $ 26.20 $       &  $ 26.92 $& \\
25   &  $ 25.39 $        &  $ 26.00 $    & $ 27.01 $       &  $ 27.33 $& \\
30   &  $ 25.46 $        &  $ 26.28 $    & $ 27.15 $       &  $ 27.25 $& \\
35   &  $ 25.33 $        &  $ 26.04 $    & $ 26.90 $       &  $ 26.97 $& \\
\enddata
\end{deluxetable}
\clearpage

\begin{deluxetable}{ccrcccccr}
\tabletypesize{\footnotesize}
\tablecaption{Detected Objects near $\epsilon$~Eri, ordered by distance from the central star.\label{objecttab}}
\tablehead{RA         &    Dec          &   dist   & size   &   {50CCD STMAG} &  notes \\
        \multispan2{(J2000; Epoch 2002.07)}    &   (")    &(pixels)&     &   }
\startdata
 3:32:56.4185 & $-$9:27:36.314  &  12.563  & 3x6  &25.2&                             \\
 3:32:56.5551 & $-$9:27:34.156  &  13.517  & 2.3  &26.7&                             \\
 3:32:56.4763 & $-$9:27:21.908  &  14.150  & 3.8  &25.4&                             \\
 3:32:56.6711 & $-$9:27:28.703  &  14.602  & 7    &25.9&                             \\
 3:32:56.8063 & $-$9:27:33.636  &  16.962  & 12   &25.7&  double object      \\
 3:32:56.8713 & $-$9:27:28.768  &  17.549  & 2.0  &26.7&                             \\
 3:32:56.8330 & $-$9:27:22.815  &  18.370  & 2.3  &26.1&                             \\
 3:32:55.9059 & $-$9:27:49.906  &  20.223  & 26   &24.2&  fuzzy patch 1.3" diam. \\
 3:32:56.5430 & $-$9:27:46.244  &  20.642  & 3x7  &25.9&  faint line                 \\
 3:32:55.4660 & $-$9:27:50.758  &  21.065  & 3.4  &25.1&                             \\
 3:32:55.7102 & $-$9:27:51.351  &  21.406  & 8x16 &23.0&  oval       \\
 3:32:56.1177 & $-$9:27:51.611  &  22.582  & 5x9  &24.4&  oval       \\
 3:32:57.1903 & $-$9:27:25.101  &  22.740  & 2.9  &26.4&                             \\
 3:32:56.5049 & $-$9:27:09.433  &  23.767  & 3.0  &25.6&               \\
 3:32:57.2179 & $-$9:27:19.614  &  24.862  & 3.2  &25.8&                             \\
 3:32:57.3883 & $-$9:27:34.610  &  25.586  & 3.0  &25.1&  extended flux              \\
 3:32:57.1796 & $-$9:27:45.611  &  27.068  & 3x10 &26.9&  faint line                 \\
 3:32:57.4928 & $-$9:27:19.772  &  28.554  & 4x6  &23.8&  oval galaxy\tablenotemark{a}  \\
 3:32:56.7705 & $-$9:27:54.110  &  28.991  & 2.3  &25.4&                             \\
 3:32:57.6676 & $-$9:27:29.800  &  29.281  & 3.4  &26.2&                             \\
 3:32:57.2929 & $-$9:27:11.811  &  29.844  & 3    &26.4&   on edge of FOV    \\
 3:32:57.8475 & $-$9:27:34.308  &  32.240  & 4.4  &24.5&                             \\
 3:32:57.2044 & $-$9:27:55.359  &  33.897  & 3.7  &26.2&                             \\
 3:32:58.0335 & $-$9:27:28.253  &  34.728  & 6.1  &26.1& extended structure mag 24.8\\
 3:32:58.0057 & $-$9:27:19.861  &  35.718  & 3.1  &26.6&                             \\
 3:32:58.0150 & $-$9:27:41.904  &  36.439  & 7.6  &25.2&                             \\
 3:32:58.1936 & $-$9:27:30.231  &  37.056  & 5x9  &25.6& extended structure   \\
 3:32:57.2856 & $-$9:27:58.997  &  37.445  & 4    &26.1&                             \\
 3:32:58.1967 & $-$9:27:36.068  &  37.606  & 4    &26.1&                             \\
 3:32:58.2965 & $-$9:27:37.911  &  39.394  & 4x10 &24.1&  long oval      \\
 3:32:58.3564 & $-$9:27:31.950  &  39.514  & 1.5  &25.6&   extended structure      \\
 3:32:58.1802 & $-$9:27:46.132  &  40.259  & 4.0  &26.1&                          \\
 3:32:58.3193 & $-$9:27:17.338  &  40.889  & 4    &27.8&                      \\
 3:32:58.0451 & $-$9:27:51.423  &  40.946  & 5    &25.5&   extended structure    \\
 3:32:58.1279 & $-$9:27:49.574  &  41.079  & 3.2  &23.2&  double irr galaxy\tablenotemark{b}\\
 3:32:58.5861 & $-$9:27:24.382  &  43.210  & 2.0  &27.4&                             \\
 3:32:56.9571 & $-$9:28:08.992  &  43.308  & 3.7  &25.3&                             \\
 3:32:58.6919 & $-$9:27:24.300  &  44.772  & 3.2  &27.1&  extended structure mag 25.8\\
 3:32:58.4244 & $-$9:27:49.437  &  44.919  & 3.8  &25.7&                             \\
 3:32:58.7317 & $-$9:27:24.345  &  45.350  & 2.5  &26.2&                             \\
 3:32:58.7489 & $-$9:27:25.229  &  45.504  & 5    &26.1&                             \\
 3:32:58.6853 & $-$9:27:19.098  &  45.620  & 11   &25.9&  noise?                     \\
 3:32:58.3840 & $-$9:27:53.969  &  46.549  & 3    &25.7&                             \\
 3:32:57.8873 & $-$9:28:04.667  &  47.568  & 2.4  &26.0&                             \\
 3:32:58.9032 & $-$9:27:31.625  &  47.574  & 2    &26.1&                             \\
 3:32:57.8779 & $-$9:28:07.866  &  49.856  & 5    &26.0&                             \\
 3:32:59.0790 & $-$9:27:37.894  &  50.771  & 5.2  &25.4&                             \\
 3:32:59.0779 & $-$9:27:38.375  &  50.833  & 3    &27.1&                             \\
 3:32:58.9754 & $-$9:27:46.790  &  51.449  & 20   &23.7&  fuzzy patch $\approx$1" diam.    \\
 3:32:58.0864 & $-$9:28:07.301  &  51.501  & 1.8  &19.8&  bright star                \\
 3:32:58.8881 & $-$9:27:52.372  &  52.366  & 3.3  &26.6&                             \\
 3:32:59.3340 & $-$9:27:29.536  &  53.912  & 1.8  &22.2&  star                       \\
 3:32:58.5123 & $-$9:28:04.604  &  54.265  & 3.3  &25.7&                             \\
 3:32:59.4291 & $-$9:27:34.188  &  55.481  &10x30 &22.6&  oval        \\
 3:32:59.3587 & $-$9:27:16.223  &  55.970  & 4    &28.3&                         \\
 3:32:59.4461 & $-$9:27:42.934  &  57.069  & 5x10 &25.6&                             \\
 3:32:59.5610 & $-$9:27:30.327  &  57.267  & 4    &26.0&                             \\
 3:32:59.5556 & $-$9:27:37.599  &  57.698  & 3.5  &23.5& oval outer isophote      \\
 3:32:59.4106 & $-$9:27:49.428  &  58.392  & 4.2  &25.1&                      
\enddata
\tablenotetext{a}{Object \#4 from Macintosh et al.\ 2003, $K=19.4$.}
\tablenotetext{b}{Object \#6 from Macintosh et al.\ 2003, $K=20.2$.}
\end{deluxetable}


\begin{thebibliography}

\bibitem[Aumann(1985)]{1985PASP...97..885A} Aumann, H.~H.\ 1985, \pasp, 97, 
885 

\bibitem[Aumann et al.(1984)]{1984ApJ...278L..23A} Aumann, H.~H.~et al.\ 
1984, \apjl, 278, L23 

\bibitem[Beichman et al.(1985)]{1985STIN...8518898B} Beichman, C.~A., 
Neugebauer, G., Habing, H.~J., Clegg, P.~E., \& Chester, T.~J.\ 1985, NASA 
STI/Recon Technical Report N, 85, 18898 

\bibitem[Burns, Lamy, \& Soter(1979)]{1979Icar...40....1B} Burns, J.~A., 
Lamy, P.~L., \& Soter, S.\ 1979, Icarus, 40, 1 

\bibitem[Chabrier, Baraffe, Allard, \& 
Hauschildt(2000)]{2000ApJ...542..464C} Chabrier, G., Baraffe, I., Allard, 
F., \& Hauschildt, P.\ 2000, \apj, 542, 464 


\bibitem[Chini, Kruegel, \& Kreysa(1990)]{1990A&A...227L...5C} Chini, R., 
Kr\"ugel, E., \& Kreysa, E.\ 1990, \aap, 227, L5 

\bibitem[Chini et al.(1991)]{1991A&A...252..220C} Chini, R., Kr\"ugel, E., 
Kreysa, E., Shustov, B., \& Tutukov, A.\ 1991, \aap, 252, 220 

\bibitem[Dohnanyi(1969)]{1969JGR....74.2431D} Dohnanyi, J.~S.\ 1969, \jgr, 
74, 2431 

\bibitem[Gillett(1986)]{1986lodm.conf...61G} Gillett, F.~C.\ 1986, ASSL 
Vol.~124: Light on Dark Matter, 61 

\bibitem[Gillett \& Aumann(1983)]{1983BAAS...15..799G} Gillett, F.~\& 
Aumann, H.~H.~G.\ 1983, \baas, 15, 788 

\bibitem[Grady et al.(2003)]{2003PASP..115.1036G} Grady, C.~A.~et al.\ 
2003, \pasp, 115, 1036 

\bibitem[Greaves et al.(1998)]{1998ApJ...506L.133G} Greaves, J.~S.~et al.\ 
1998, \apjl, 506, L133 

\bibitem[Hatzes et al.(2000)]{2000ApJ...544L.145H} Hatzes, A.~P.~et al.\ 
2000, \apjl, 544, L145 

\bibitem[Holland et al.(1998)]{1998Natur.392..788H} Holland, W.~S.~et al.\ 
1998, \nat, 392, 788 

\bibitem[Kim Quijano et al.(2003)]{STISIHB}
Kim Quijano, J., et al. 2003, "STIS Instrument Handbook", Version 7.0, (Baltimore: STScI).

\bibitem[Krist(1993)]{1993adass...2..536K} Krist, J.\ 1993, ASP Conf.~Ser.~ 
52: Astronomical Data Analysis Software and Systems II, 2, 536 

\bibitem[Krist(1995)]{1995adass...4..349K} Krist, J.\ 1995, ASP Conf.~Ser.~ 
77: Astronomical Data Analysis Software and Systems IV, 4, 349 

\bibitem[Kruegel(2003)]{2003pid..book.....K} Kr\"ugel, E.\ 2003, The
physics  of interstellar dust, by Endrik Kr\"ugel.~IoP Series in
astronomy and  astrophysics, ISBN 0750308613.~Bristol, UK: The Institute
of Physics

\bibitem[Li \& Greenberg(1997)]{1997A&A...323..566L} Li, A.~\& Greenberg, 
J.~M.\ 1997, \aap, 323, 566 

\bibitem[Li \& Greenberg(1998)]{1998A&A...331..291L} Li, A.~\& Greenberg, 
J.~M.\ 1998, \aap, 331, 291 

\bibitem[Li \& Lunine(2003a)]{2003ApJ...590..368L} Li, A.~\& Lunine, J.~I.\ 
2003a, \apj, 590, 368 

\bibitem[Li \& Lunine(2003b)]{2003ApJ...594..987L} Li, A.~\& Lunine, J.~I.\ 
2003b, \apj, 594, 987 

\bibitem[Li, Lunine, \& Bendo(2003)]{2003ApJ...598L..51L} Li, A., Lunine, 
J.~I., \& Bendo, G.~J.\ 2003, \apjl, 598, L51 

\bibitem[Liou \& Zook(1999)]{1999AJ....118..580L} Liou, J.~\& Zook, H.~A.\ 
1999, \aj, 118, 580 

\bibitem[Lisse et al.(1998)]{1998ApJ...496..971L} Lisse, C.~M., A'Hearn, 
M.~F., Hauser, M.~G., Kelsall, T., Lien, D.~J., Moseley, S.~H., Reach, 
W.~T., \& Silverberg, R.~F.\ 1998, \apj, 496, 971 

\bibitem[Macintosh et al.(2003)]{2003ApJ...594..538M} Macintosh, B.~A., 
Becklin, E.~E., Kaisler, D., Konopacky, Q., \& Zuckerman, B.\ 2003, \apj, 
594, 538 

\bibitem[Mermilliod, Mermilliod, \& Hauck(1997)]{1997A&AS..124..349M} 
Mermilliod, J.-C., Mermilliod, M., \& Hauck, B.\ 1997, \aaps, 124, 349 

\bibitem[Moro-Mart{\'{\i}}n \& Malhotra(2002)]{2002AJ....124.2305M} 
Moro-Mart{\'{\i}}n, A.~\& Malhotra, R.\ 2002, \aj, 124, 2305 

\bibitem[Moro-Mart{\'{\i}}n \& Malhotra(2003)]{2003AJ....125.2255M} 
Moro-Mart{\'{\i}}n, A.~\& Malhotra, R.\ 2003, \aj, 125, 2255 

\bibitem[Ozernoy, Gorkavyi, Mather, \& Taidakova(2000)]{2000ApJ...537L.147O}
Ozernoy, L.~M., Gorkavyi, N.~N.,  Mather, J.~C., \& Taidakova, T.~A.\ 2000, \apjl,
537, L147 

\bibitem[Pavlovsky, C., et al.(2003)]{acs_ihb_c13} Pavlovsky, C., et al.\
2003, ``ACS Instrument Handbook'', Version 4.0, (Baltimore: STScI).

\bibitem[Sch{\" u}tz et al.(2004)]{2004A&A...414L...9S} Sch{\" u}tz, O., 
Nielbock, M., Wolf, S., Henning, T., \& Els, S.\ 2004, \aap, 414, L9 

\bibitem[Sheret, Dent, \& Wyatt(2003)]{2003astro.ph.11593S} Sheret, I., 
Dent, W.~R.~F., \& Wyatt, M.~C.\ 2003, ArXiv Astrophysics e-prints, 11593 

\bibitem[Soderblom \& Dappen(1989)]{1989ApJ...342..945S} Soderblom, 
D.~R.~\& Dappen, W.\ 1989, \apj, 342, 945 


\bibitem[Song et al.(2000)]{2000ApJ...533L..41S} Song, I., Caillault, 
J.-P., Barrado y Navascu{\' e}s, D., Stauffer, J.~R., \& Randich, S.\ 2000, 
\apjl, 533, L41 

\bibitem[Tanaka, Inaba, \& Nakazawa(1996)]{1996Icar..123..450T} Tanaka, H., 
Inaba, S., \& Nakazawa, K.\ 1996, Icarus, 123, 450 

\bibitem[Quillen \& Thorndike(2002)]{2002ApJ...578L.149Q} Quillen, A.~C.~\& 
Thorndike, S.\ 2002, \apjl, 578, L149 

\bibitem[Weintraub \& Stern(1994)]{1994AJ....108..701W} Weintraub, D.~A.~\& 
Stern, S.~A.\ 1994, \aj, 108, 701 

\bibitem[Wiscombe(1979)]{1979msca.rept.....W} Wiscombe, W.~J.\ 1979,    
NCAR Tech Note TN-140+STR, National Center For
Atmospheric Research, Boulder, Colorado.

\bibitem[Wiscombe(1980)]{1980ApOpt..19.1505W} Wiscombe, W.~J.\ 1980, \ao, 
19, 1505 

\bibitem[Wyatt \& Dent(2002)]{2002MNRAS.334..589W} Wyatt, M.~C.~\& Dent, 
W.~R.~F.\ 2002, \mnras, 334, 589 

\bibitem[Wyatt et al.(1999)]{1999ApJ...527..918W} Wyatt, M.~C., Dermott, 
S.~F., Telesco, C.~M., Fisher, R.~S., Grogan, K., Holmes, E.~K., \& Pi{\~ 
n}a, R.~K.\ 1999, \apj, 527, 918 

\bibitem[Zuckerman \& Becklin(1993)]{1993ApJ...414..793Z} Zuckerman, B.~\& 
Becklin, E.~E.\ 1993, \apj, 414, 793 

\end{thebibliography}
\end{document}